\documentclass[12pt]{article}

\usepackage{times}
\usepackage{epsfig}
\usepackage{amssymb}
\usepackage{amsfonts}
\usepackage{amsmath}
\usepackage{amssymb}
\usepackage{amstext}
\usepackage{amsthm}
\usepackage{subfigure}	
\usepackage{graphicx}
\usepackage{color}								
\usepackage{jheppub} 
\begin{document}

\newcommand{\nc}{\newcommand}
\newcommand{\rnc}{\renewcommand}

\rnc{\baselinestretch}{1.24}    
\setlength{\jot}{6pt}       
\rnc{\arraystretch}{1.24}   

\makeatletter
\rnc{\theequation}{\thesection.\arabic{equation}}
\@addtoreset{equation}{section}
\makeatother


 \renewcommand{\thefootnote}{\fnsymbol{footnote}}
                                                   
\newcommand{\tcr}{\textcolor{red}}
\newcommand{\tcg}{\textcolor{green}}           
\newcommand{\tcb}{\textcolor{blue}}


\def\be{\begin{eqnarray}}
\def\ee{\end{eqnarray}}
\def\nn{\nonumber\\}


\def\ct{\cite}
\def\la{\label}
\def\eq#1{\eqref{#1}}


\def\a{\alpha}
\def\b{\beta}
\def\g{\gamma}
\def\G{\Gamma}
\def\d{\delta}
\def\D{\Delta}
\def\e{\epsilon}
\def\et{\eta}
\def\ph{\phi}
\def\Ph{\Phi}
\def\ps{\psi}
\def\Ps{\Psi}
\def\k{\kappa}
\def\l{\lambda}
\def\L{\Lambda}
\def\m{\mu}
\def\n{\nu}
\def\th{\theta}
\def\Th{\Theta}
\def\r{\rho}
\def\s{\sigma}
\def\S{\Sigma}
\def\ta{\tau}
\def\o{\omega}
\def\O{\Omega}
\def\pr{\prime}


\def\half{\frac{1}{2}}
\def\goto{\rightarrow}

\def\na{\nabla}
\def\grad{\nabla}
\def\curl{\nabla\times}
\def\div{\nabla\cdot}
\def\pa{\partial}
\def\fr{\frac}

\def\bra{\left\langle}
\def\ket{\right\rangle}
\def\lb{\left[}
\def\lc{\left\{}
\def\ls{\left(}
\def\lp{\left.}
\def\rp{\right.}
\def\rb{\right]}
\def\rc{\right\}}
\def\rs{\right)}

\def\vac#1{\mid #1 \rangle}


\def\td#1{\tilde{#1}}
\def\check{ \maltese {\bf Check!}}


\def\Tr{{\rm Tr}\,}
\def\det{{\rm det}}
\def\text#1{{\rm #1}}


\def\bc#1{\nnindent {\bf $\bullet$ #1} \\ }
\def\ch {$<Check!>$ }
\def\ss {\vspace{1.5cm}}
\def\inf{\infty}

\begin{titlepage}

\hfill\parbox{2cm} { }

%
\vspace{1cm}

\begin{center}
{\Large \bf  Holographic two-point functions in a disorder system}

\vskip 1. cm
   {Chanyong Park$^{a}$\footnote{e-mail : cyong21@gist.ac.kr}}

\vskip 0.5cm

{\it $^a$ Department of Physics and Photon Science, Gwangju Institute of Science and Technology, \\ 
Gwangju  61005, Korea}

\end{center}

\thispagestyle{empty}

\vskip2cm


\centerline{\bf ABSTRACT} \vskip 4mm

\vspace{1cm}

We study the holographic dual of two-point correlation functions for nonconformal field theories. We first take into account a Lifshitz geometry as the dual of a Lifshitz field theory which may appear at a critical or IR fixed point. We explicitly show the holographic relation between a Lifshitz geometry and a Lifshitz field theory by calculating  two-point correlators and equation of state parameter on both sides. We also investigate a disorder deformation, which allows a UV conformal field theory to flow into a new IR Lifshitz field theory. In this deformed theory, we investigate an anomalous dimension representing the change of an operator's scaling dimension along the RG flow.

\vspace{2cm}

\end{titlepage}

\renewcommand{\thefootnote}{\arabic{footnote}}
\setcounter{footnote}{0}



\section{Introduction}

Knowing correlation functions is important to understand the nontrivial quantum nature of strongly interacting systems and their physical properties relying on an energy scale. Despite this fact, it is still hard to calculate correlators nonperturbatively due to the absence of a nonperturbative calculation technique. In this situation, the AdS/CFT correspondence recently proposed that nonperturbative features of quantum field theories (QFT) can be captured by classical gravity theories \cite{Maldacena:1997re,Gubser:1998bc,Witten:1998qj,Witten:1998zw}. Therefore, it would be interesting to investigate how to describe nonperturbative two-point functions holographically. 

According to the AdS/CFT correspondence, a $d$-dimensional strongly interacting conformal field theory (CFT) maps to a gravity theory in a $(d+1)$-dimensional AdS space. Due to the large symmetry of a CFT and AdS space, various nonperturbative quantities were well studied on the dual gravity side. For example, conformal symmetry determines two-point functions up to normalization. Intriguingly, it was shown that the holographic renormalization can reproduce the same two-point functions exactly \cite{Gubser:1998bc,Witten:1998qj,Witten:1998zw}. These results are robust because of the large symmetry. However, when we take into account a nonconformal QFT, can we calculate its nonperturbative correlators holographically? If there is no sufficient large symmetry, it is not easy to find exact correlation functions not only in the QFT but also in the holographic setup.

Another interesting feature of the AdS/CFT correspondence is that some physical quantities of a QFT can be realized as geometrical objects on the dual gravity side. For example, it was shown that the $q\bar{q}$-potential \cite{Maldacena:1998im,Rey:1998bq,Park:2009nb} and entanglement entropy \cite{Ryu:2006bv,Ryu:2006ef,Myers:2012ed,Blanco:2013joa,Wong:2013gua,Bhattacharya:2012mi,Momeni:2015vka,Fischler:2012ca,Kim:2016jwu,Park:2015dia} can be described by a string worldsheet and minimal surface extending to the dual geometry. Similarly, it was also conjectured that a two-point function maps to a geodesic curve on the dual gravity side \cite{Susskind:1998dq,Solodukhin:1998ec,DHoker:1998vkc,Liu:1998ty,Balasubramanian:1999zv,Louko:2000tp,Kraus:2002iv,Fidkowski:2003nf,Park:2020nvo,Rodriguez-Gomez:2021pfh}
\be			\la{Formula:twopoint}
\bra O (t_1,x_1) O(t_2,x_2) \ket \sim e^{- \D L(t_1,x_1;t_2, x_2) /R} ,
\ee
where $\D$ is the conformal dimension of an operator $O$ and $ L(t_1,x_1;t_2, x_2)$ indicates a geodesic length connecting two boundary operators. Applying this proposal to an AdS space, one can easily reproduce the CFT's two-point function. Although there is no rigorous proof on this alternative method, it is still fascinating in that it can easily generalize to a nonconformal QFT and non-AdS geometry.

In order to check the validity of \eq{Formula:twopoint}, we first study two-point functions of a Lifshitz field theory (LFT). A LFT is not a CFT because it breaks the Lorentz symmetry. Therefore, temporal and spatial coordinates of a LFT behave in different ways. Due to the absence of the Lorentz symmetry, we need to introduce two different two-point functions, temporal (equal-position or auto-correlation) and spatial (equal-time) correlation functions. In this case, temporal or spatial two-point functions indicates correlation between two operators whose distance is time-like or space-like, respectively. However, a LFT still has a scaling symmetry which enables us to calculate two-point functions analytically. In the holographic study, a Lifshitz geometry is believed as the dual of a LFT because they have the same scaling symmetry \cite{Balasubramanian:2008dm,Goldberger:2008vg,Adams:2008wt,Maldacena:2008wh,Minic:2008xa,Kovtun:2008qy,Lee:2008xf,Lin:2008pi,Mazzucato:2008tr,Rangamani:2008gi,Akhavan:2008ep,Kachru:2008yh,Taylor:2008tg,Keranen:2012mx,Keranen:2016ija,KordZangeneh:2017zgg,Park:2013goa,Park:2013dqa,Park:2014raa}. In this work, we calculate temporal and spatial two-point functions of a LFT and then compare them with the holographic results derived in the dual Lifshitz geometry. We find that the LFT's results perfectly match to those of the dual Lifshitz geometry. Moreover, we calculate the equation of state parameter of the LFT's excitation. We show that the same equation of state parameter appears in the Lifshitz black hole. This indicates that the Lifshitz black hole can be regarded as thermalization of the LFT excitation.

A LFT may appear at a critical or IR fixed point with restoring the scaling symmetry. For example, when we deform a CFT by a relevant operator, a UV CFT can flow into another IR theory along the renormalization group (RG) flow \cite{park:2021nyc}. When a CFT deforms by a disorder, for example, its holographic dual gravity has been known in Ref. \cite{Hartnoll:2014cua,Narayanan:2018ilr}. In this case, a scaling dimension is well defined only at fixed points where the scaling symmetry is restored. At the UV and IR fixed points, in general, an operator can have different scaling dimensions. To describe the change of a scaling dimension, we define an effective scaling dimension depending on the RG scale. In the holographic dual of a disordered system, we investigate how the disorder modifies an effective scaling dimension along the RG flow.

The rest of this paper is as follows. In Sec. 2, we first discuss a simple LFT and its holographic dual. In Sec. 3, we holographically investigate two-point functions of the LFT vacuum and thermalized LFT. In Sec. 4, we consider a relevant disorder which deforms a UV CFT into an IR LFT.  We study how the disorder modifies the scaling dimension of an operator along the RG flow. In Sec. 5, we finish this work with some concluding remarks.


\section{Holographic dual of Lifshitz field theory}

Let us take into account a simple LFT in a $d$-dimensional Euclidean space  \cite{Kachru:2008yh,Taylor:2008tg,Keranen:2012mx,Keranen:2016ija}
\be
S_{LFT} \equiv \int d \ta \, d^{d-1} x \,  {\cal L} 
= \int d \ta \, d^{d-1} x \, \sqrt{g} \lb  \fr{1}{2}  \, ( \pa_\ta \ph )^2   +  \fr{1}{2}   \ls  \pa^\l \ph \rs^2 + V(\ph)  \rb  , 
\ee
where $\pa = \sqrt{\pa^i \pa_i}$ with $i=1,\cdots,d-1$ and $g_{\ta\ta} = g_{ii}=1$. We first assume that a dynamical critical exponent $\l$ is an integer and then generalize it into a real number by analytic continuation. When a scalar potential is absent, the above LFT reduces to a free theory with a scaling symmetry \cite{Taylor:2008tg,Park:2013goa,Park:2013dqa}
\be     \la{Transform:scale}
\ta \to \d^{-\l} \ta \ , \quad x_i \to \d^{-1} x_i  \quad {\rm and} \quad  \ph  \to \d^{\D_\ph}  \ph ,
\ee
where the scaling dimension of $\ph$ is given by $\D_\ph = (d-1-\l)/2$. Since temporal and spatial coordinates transform differently, the LFT generally breaks a Lorentz symmetry except for $\l=1$. For $\l=1$, in particular, the LFT lifts up to a CFT with restoring a conformal symmetry.

Above a general scalar potential spoils the scaling symmetry. However, the following specific scalar potential still preserves the scaling symmetry  
\be
V (\ph) = \k \  \ph^{2(d-1+\l)/(d-1-\l)} ,
\ee
where $\k$ is a dimensionless coupling constant. From the free LFT viewpoint, this scalar potential corresponds to a marginal deformation with a scaling dimension 
\be
\D_{m} =   d - 1 + \l  .   \la{Result:marginaldim}
\ee 
For $\l=1$, $\D_m$ reduces to $\D_m=d$ which is the conformal dimension of a CFT's marginal operator. For the LFT, an operator in the range of $0 < \D < \D_{m}$ corresponds to a relevant one, while it becomes irrelevant for $\D > \D_{m}$.

For the free LFT, variation of the action leads to the following equation of motion
\be
0 =  \lb - \pa_\ta^2 +  (-1)^\l   \pa^{2 \l} \rb \ph .
\ee
Under the Fourier transformation
\be
\ph  (t,x^i) =  \int d \o d^{d-1} p \ {\varphi} (\o, p_i)  \, e^{-i (\o t + p_i x^i)}  ,   \la{Result:dreli}
\ee
the solution of the equation of motion must satisfy the following dispersion relation
\be
0 = w^2 +  p^{2 \l} ,  \la{Result:drel}
\ee
where the momentum $p$ is defined as $p^2 = \sum_{i=1}^{d-1} p_i^2$. Now, we take into account an energy-momentum tensor of this system. For the Euclidean LFT, 
the energy and pressure are give by
\be
{T^{\ta}}_{\ta} &=& 
 \ls \pa_\ta \ph \rs^2 - \fr{1}{2} \lb  \ls \pa_\ta \ph \rs^2 + \sum_{i=1}^{d-1}  \ls \pa_i^\l \ph \rs^2 \rb  , \\
{T^i}_{i} &=& 
 \l \ls \pa_i^\l \ph \rs^2 - \fr{1}{2} \lb  \ls \pa_\ta \ph \rs^2 + \sum_{i=1}^{d-1}  \ls \pa_i^\l \ph \rs^2 \rb   .
\ee

After the inverse Wick rotation, the energy and pressure in the Minkowski spacetime become
\be
E = - \int d^{d-1} x \ {T^{\ta}}_{\ta}  \quad {\rm and} \quad
P V = \int d^{d-1} x \ {T^i}_{i}   \la{Result:LFTEM}
\ee
where  $V = \int d^{d-1} x $ is an appropriately regularized volume. These quantities together with the dispersion relation \eq{Result:drel} yield the following equation of state parameter 
\be
w = \fr{P V}{E}  
= \fr{\l}{d-1} ,  \la{Result:EoSparameter}
\ee
and the trace of the energy-momentum tensor reads
\be
{T^\m}_\m = (\l -1) E .
\ee
For $\l=1$, the energy-momentum becomes traceless and the LFT  lifts up to a CFT, as mentioned before. Recalling that the vacuum energy of the free LFT vanishes, \eq{Result:LFTEM} and \eq{Result:EoSparameter} represent the energy-momentum tensor and equation of state parameter of the LFT's excitation. For $\l=1$, \eq{Result:EoSparameter} reduces to $w=1/(d-1)$ which describes massless excitation of a CFT \cite{Park:2012lzs}.

For the free LFT, a two-point function appears as a solution of the following equation
\be
\lb -  \pa_\ta^2 +  (-1)^\l  \pa^{2 \l}\rb   G(x,x')  = \d^{(d)} (x-x')   .
\ee
In the momentum space, therefore, the two-point function reads
\be				\la{Result:towptfninmom}
G(\o, p)  =   \fr{1}{\o^2 +  p^{2 \l}} .
\ee
Using the inverse Fourier transformation, the two-point function in the position space is given by
\be
\bra \ph (\ta_1, \vec{x}_1) \,  \ph (\ta_2,\vec{x}_2) \ket   
&=&  \O_{d-3} \int d \o \,  d p \, d \th  \ p^{d-2} \sin^{d-3} \th  \ G(\o,p)  \ e^{ 
	i \ls \o (  \ta_1 - \ta_2  ) - p | \vec{x}_1-\vec{x}_2  | \cos \th  \rs }   ,
\ee
where $ \O_{d-3} $ indicates a solid angle of a $(d-3)$-dimensional unit sphere. Due to the absence of a Lorentz symmetry, LFT's two-point functions usually have different forms in the temporal and spatial directions. 

Let us first take into account a spatial two-point function. When the distance of two operators is space-like, they are causally disconnected in the Minkowski spacetime. Therefore, there is no classical correlation. At the quantum level, however, non-locality of a quantum theory allows nontrivial correlation. As a result, a nonvanishing spatial two-point function represents quantum correlation between two causally disconnected operators. Assuming that two local operators are sitting at an equal time $\ta$ and different positions, a spatial two-point function reads up to normalization
\be
\bra \ph (\ta, \vec{x}_1) \,  \ph (\ta, \vec{x}_2) \ket   
&=& \O_{d-3} \int d \o \,  d p\, d \th  \ p^{d-2} \sin^{d-3} \th  \ G(\o,\vec{p})  \ e^{ 
	- i  p  | \vec{x}_1-\vec{x}_2  | \cos \th } 
\sim  \fr{1}{|\vec{x}_1-\vec{x}_2|^{2 \D_\ph}}   .   \quad
\ee
Using the Wick contraction, the spatial two-point function of a general operator, $O=\ph^n$, yields
\be
\bra O (\ta, \vec{x}_1) \,  O ( \ta, \vec{x}_2) \ket \sim \fr{1}{|\vec{x}_1-\vec{x}_2|^{2 \D}} ,   \la{Result:spatial2}
\ee
where $\D = n \D_\ph$. This is the same as the CFT's result because the LFT considered here is different from a CFT only in the temporal direction.

We move to a temporal two-point function which represents correlation between two operators whose distance is time-like. Here, a time-like distance implies that two operators are causally connected. Therefore, a temporal two-point function describes time evolution of a local operator. When an operator evolves at a fixed position $\vec{x}$, a temporal two-point function becomes
\be
\bra \ph (\ta_1, \vec{x}) \,  \ph ( \ta_2, \vec{x}) \ket   
&=& \O_{d-3} \int d \o \,  d p \, d \th  \ p^{d-2} \sin^{d-3} \th  \ G(\o,\vec{p})  \ e^{ 
	i  \o (\ta_1 - \ta_2) }  
\sim  \fr{1}{| \ta_1 - \ta_2 |^{2 \D_{\ph}/\l}}   .
\ee
For a general operator $O$ with a scaling dimension $\D$, it further generalizes into 
\be
\bra O ( \ta_1, \vec{x}) \,  O (\ta_2, \vec{x}) \ket   &\sim& \fr{1}{| \ta_1 - \ta_2 |^{2 \D /\l}}  .  \la{Result:ttwopLFT}
\ee
The results in  \eq{Result:spatial2} and \eq{Result:ttwopLFT} are typical two-point functions of a LFT \cite{Keranen:2012mx,Keranen:2016ija}.


Now, let us study the holographic dual of the above  LFT. To do so, we take into account a Euclidean Lifshitz geometry which is obtained from the following Einstein-Maxwell-scalar theory with a negative cosmological constant $\L$ \cite{Taylor:2008tg,Park:2013goa,Park:2013dqa,Park:2014raa}
\be
S_{gr} = \fr{1}{2 \k^2} \int d^{d+1} x \sqrt{g} \ls  {\cal R} + 2 \L + \half \pa_\m \ph \pa^\m \ph + \fr{1}{4} 
e^{\a \ph} F_{\m\n} F^{\m\n}  \rs ,
\ee
where $\a$ is an appropriate constant determining a critical dynamical exponent. Assuming that the gauge field has only a time component to break a Lorentz symmetry, the above Einstein-Maxwell-scalar theory allows a Lifshitz geometry as a solution 
\be
ds^2 = \fr{R^2}{z^2}   \ls  \fr{R^{2(\l-1)}}{z^{2(\l-1)}} d \ta^2 + d \vec{x}^2_{d-1} + d z^2\rs .
\ee
For $d=4$, a dynamical critical exponent $\l$ is related to the intrinsic parameter $\a$ \cite{Park:2013goa,Park:2013dqa,Park:2014raa}
\be
\a = \fr{2}{\sqrt{\l -1}} .
\ee
The Lifshitz metric is invariant under the following transformation
\be
\ta \to \d^{- \l} \ta \ , \quad \vec{x} \to \d^{-1} \vec{x} \quad {\rm and } \quad z \to \d^{-1} z . 
\ee
This is equivalent to \eq{Transform:scale}, which makes us believe the Lifshitz geometry as the dual of a LFT.

Let us consider a scalar field fluctuation $\varphi$ on the Lifshitz geometry. Assuming that $\varphi$ relies only on the radial coordinate, its equation of motion reads
\be
0
&=& \pa_z^2 \varphi -  \fr{d+\l -2}{z} \pa_z  \varphi - \fr{m^2}{z^2} \varphi .
\ee
where $m$ indicates a mass of $\varphi$.  In an asymptotic region ($z \to 0$), a perturbative solution is given by
\be
\varphi  = c_1\  z^{\D_-}  + c_2 \ z^{\D_+} ,
\ee
with
\be
\D_{\pm} = \fr{d-1+\l}{2} \pm \half \sqrt{ (d-1+\l)^2 + 4 m^2} .
\ee
According to the AdS/CFT correspondence for $\l=1$, $c_1$ and $c_2$ are identified with a coupling constant (or source) and the vev of a dual operator, respectively. We also apply the similar identification to the LFT. Although the LFT has no conformal symmetry, $\D_+$ is still identified with a scaling dimension because of the scaling symmetry of the LFT. For $m=0$, the coupling constant becomes dimensionless and the dual operator is marginal. Therefore, the scaling dimension of a marginal operator is given by
\be
\D_{m} = d-1 + \l ,
\ee  
which is exactly the same as \eq{Result:marginaldim} obtained in the LFT. When $m^2$ is in the range of $-(d-1+\l)^2 /4 < m^2 < 0$, the scalar field rapidly suppresses in the asymptotic region. Therefore, the dual operator becomes relevant. For $m^2 > 0$, the dual operator is irrelevant and the source term does not vanish in the asymptotic region. This indicates that the gravitational backreaction of the scalar field seriously modifies the asymptotic geometry. This is the typical feature of an irrelevant operator.

The above Lifshitz geometry can further generalize into a Lifshitz black hole 
\be         \la{Metric:deformedLif}
ds^2 = \fr{R^2}{z^2}   \ls \fr{R^{2(\l-1)}}{z^{2(\l-1)}}  f(z) \, d \ta^2 + d \vec{x}^2_{d-1} + \fr{dz^2}{f(z)}\rs ,
\ee
with a blackening factor
\be
f(z) = 1 - \ls \fr{z}{z_h} \rs^{d-1+ \l}  ,
\ee
where $z_h$ indicates a black hole horizon. According to the AdS/CFT correspondence, the Lifshitz black hole maps to the LFT at finite temperature. In other words, the Lifshitz geometry is the dual of the LFT vacuum, whereas a Lifshitz black hole describes thermalization of the LFT excitation \cite{Park:2013dqa,Park:2013goa,Park:2014raa}. To see this feature, we look into thermodynamics of the Lifshitz black hole. For a $(d+1)$-dimensional Lifshitz black hole, the Hawking temperature and Bekenstein-Hawking entropy are given by
\be
T_H = \frac{(d+\l-1)}{4 \pi }  \fr{1}{ z_h^{\l}} \quad {\rm and} \quad
S_{BH} = \frac{ R^{d-1} V }{4 G}  \fr{1}{ {z_h}^{d-1}}.
\ee
The first law of thermodynamics determines an internal energy and pressure as  
\be
E =  \frac{(d-1)  R ^{d-1} V }{16 \pi  G}   \fr{1}{ z_h^{d+\l-1}}  ,  
\quad {\rm and} \quad  P = \frac{\l R^{d-1} }{16 \pi  G}   \fr{1}{z_h^{d+\l-1}} .  \la{Result:exenergy}
\ee
These quantities satisfy the equation of state parameter in \eq{Result:EoSparameter}. This implies that the Lifshitz black hole is dual to thermalization of the LFT excitation. Lastly, a heat capacity at a given volume reads
\be
c_V \equiv \fr{T_H dS_{BH}}{d T_H}= \frac{(d-1) R^{d-1} V }{4 G \l}   \fr{1}{ z _h^{d-1}}   \ge 0.
\ee
Here, the non-negative heat capacity means that the dual LFT is thermodynamically stable.

\section{Holographic two-point functions }

\subsection{Holographic two-point functions in the LFT vacuum}

We study holographic two-point functions in the Lifshitz geometry which are the dual of two-point functions in the LFT vacuum. When a local operator $O (\ta,x)$ has a scaling dimension $\D$, its holographic two-point function is determined by \cite{Susskind:1998dq,Solodukhin:1998ec,DHoker:1998vkc,Liu:1998ty,Balasubramanian:1999zv,Louko:2000tp,Kraus:2002iv,Fidkowski:2003nf,Park:2020nvo,Rodriguez-Gomez:2021pfh}
\be    	
\bra O (\ta_1,\vec{x}_1) \,  O (\ta_2,\vec{x}_2) \ket \sim e^{-\D  \, L \ls \ta_1, \vec{x}_1 ; \ta_2 ,\vec{x}_2 \rs \, /R} ,
\ee
where $L \ls \ta_1, \vec{x}_1 ; \ta_2 ,\vec{x}_2  \rs $ is a geodesic length connecting two local operators. For a spatial two-pont function, the rotational symmetry enables us to rearrange two operator's positions to be at $\vec{x}_1=\lc 0, 0, \cdots, 0\rc$ and $\vec{x}_2=\lc   | \vec{x}_1-\vec{x}_2|  , 0, \cdots, 0 \rc$. Then, the geodesic length at a given time $\ta$ is determined by 
\be
L( \ta , \vec{x}_1 ; \ta ,\vec{x}_2 ) = R \int_0^{| \vec{x}_1-\vec{x}_2| } d x  \ \fr{ \sqrt{1 + z'^2}}{z} ,
\ee
where the prime means a derivative with respect to $x$. This up to a UV divergence results in 
\be
L(\ta, \vec{x}_1 ; \ta ,\vec{x}_2) = 2 R \log |\vec{x}_1-\vec{x}_2| ,
\ee
so that the corresponding spatial two-point function is finally given by
\be
\bra  O (\ta, \vec{x}_1) \,  O (\ta, \vec{x}_2)  \ket  \sim  \fr{1}{|\vec{x}_1-\vec{x}_2 |^{2 \D}}  .  \la{Result:Stowfun}
\ee
This is the spatial two-point function \eq{Result:spatial2} of the LFT vacuum.

Let us move to a temporal two-point function in the Euclidean Lifshitz geometry. To do so, we take into account a geodesic connecting two operators located at the same position and different time. Then, a temporal two-point function is determined by
\be
L( \ta_1, \vec{x} ; \ta_2 ,\vec{x} ) =  \int_0^{| \ta_1-\ta_2| } d \ta  \  \fr{R}{z} \  \sqrt{\fr{R^{2(\l-1)}}{z^{2(\l-1)}} + \dot{z}^2} ,
\ee
where the dot indicates a derivative with respect to $\ta$. The conserved energy of this system is given by
\be
H = -\frac{R^{2 \lambda -1}}{z^{2 \lambda -1} \sqrt{ \dot{z}^2+R^{2 ( \lambda -1) }/ z^{2 ( \lambda-1) }}} .
\ee
Denoting a turning point as $z_0$ where $\dot{z}$ vanishes, we find
\be
\dot{z}= \fr{R^{\lambda -1} \sqrt{{z_0}^{2 \lambda }-z^{2 \lambda }}}{z^{2 \lambda -1 } }  .
\ee
Using this relation, the geodesic length results in  
\be
L(\ta_1, \vec{x} ; \ta_2 ,\vec{x}  )  = \fr{2R}{\l} \log \ls \fr{ \l \, R^{\l-1} \, | \ta_1-\ta_2|}{\e^\l} \rs  ,
\ee
which determines a temporal two-point function up to normalization \cite{Keranen:2012mx,Keranen:2016ija}
\be    	
\bra O (\ta_1, \vec{x} ) \,  O (\ta_2, \vec{x} ) \ket   \sim  \fr{1}{| \ta_1- \ta_2 |^{2 \D/\l}} .   \la{Result:Liftemptwo}
\ee
This is the temporal two-point function \eq{Result:ttwopLFT} expected in the LFT vacuum.  

\subsection{ Holographic two-point functions in thermal LFT}

We further take into account an operator interacting with the LFT excitation.  Since the Lifshitz  black hole geometry corresponds to a thermal state involving all LFT excitation, the two-point function evaluated in a Lifshitz black hole describes correlation of a local operator interacting with the LFT excitation. First, we look into a spatial two-point function. In the Lifshitz black hole, a geodesic length connecting two boundary points, $\vec{x}_1$ and $\vec{x}_2$, is given by
\be
L ( \ta , \vec{x}_1 ; \ta ,\vec{x}_2 ) = \int_0^{ | \vec{x}_1-\vec{x}_2| } dx  \ \fr{R}{z} \sqrt{1 + \fr{z'^2}{f} } ,
\ee
For $f=1$, we reobtain the result of the LFT vacuum. Since the geodesic length depends on $x$ implicitly, there exists a conserved quantity
\be
H = - \fr{R \sqrt{f}}{z \sqrt{f + z'^2}}.
\ee
In addition, the invariance of the geodesic length under $x \to |\vec{x}_1 - \vec{x}_2 | - x$ allows a turning point at $x_0 = |\vec{x}_1 - \vec{x}_2 | /2$. Denoting the position of a turning point in the $z$-direction as $z_0$, $z'$ becomes zero at $z=z_0$. At the turning point, the conserved quantity further reduces to $H = -  R/z_0$. Comparing these two conserved quantities, the operator's distance at the boundary and geodesic length are determined by 
\be
|\vec{x}_1 - \vec{x}_2 | &=&  \int_0^{z_0} dz \fr{ 2 z}{\sqrt{f} \, \sqrt{z_0^2 - z^2}} ,   \la{Formula:dist} \\
L(\ta , \vec{x}_1 ; \ta ,\vec{x}_2 ) &=& \int_0^{z_0} dz  \fr{2 R z_0}{z \sqrt{f} \, \sqrt{z_0^2 - z^2}} .  \la{Formula:Glength}
\ee 

For $z_0/z_h \ll1$, two operators have a short distance. In this short-distance or UV limit, the spatial two-point function becomes
\be
\bra O ( \ta, \vec{x}_1) O( \ta, \vec{x}_2) \ket  
\approx \fr{1 } {| \vec{x}_1-\vec{x}_2|^{2 \D} } \ e^{- ( | \vec{x}_1-\vec{x}_2 | /\xi_{uv} )^{  d+\l-1} } ,
\ee  
where a UV correlation length is proportional to the energy of the LFT excitation in \eq{Result:exenergy}
\be
 \fr{1}{\xi_{uv}^{d+\l-1}}  =   \frac{\Delta  \pi ^{3/2}   R^{1-d} G  \, \Gamma   \left(\frac{ d+\lambda -1}{2}\right)}{  (d-1)  \, 2^{d+\lambda -3}  \,  \Gamma  \left(\frac{d+\lambda +2}{2} \right)}  \ \fr{E}{V} .
\ee
When the correlation length diverges ($E \to 0$), a scaling symmetry is restored and the two-point function reduces to that of the LFT vacuum. This shows that the two-point function at a given distance decreases with increasing the excitation energy due to the screening effect of the LFT excitation.

In an IR region ($z_0 \to z_h$), the operator's distance in \eq{Formula:dist} diverges at $z_0 = z_h$, so that the IR region corresponds to a long-distance limit. In the long-distance limit  ($|\vec{x}_1 - \vec{x}_2 |  \to \infty$), the geodesic length in \eq{Formula:Glength} can be rewritten as
\be
L(\ta , \vec{x}_1 ; \ta ,\vec{x}_2 )  =  \lim_{z_0 \to z_h} \ls \fr{R \ |\vec{x}_1 - \vec{x}_2 | }{z_0 } + \fr{2 R}{z_0} \int_\e^{z_0} dz \fr{\sqrt{z_0^2 -z^2}}{z \sqrt{f}}  \rs ,
\ee
where $\e$ is a UV cutoff. Here, the second term includes a logarithmic divergence at $\e \to 0$ which corresponds to a UV divergence. After renormalizing the UV divergence, the second term becomes finite even at $z_0 =z_h$. This fact indicates that the first term is dominant in the long-distance limit. Therefore, the leading behavior of the spatial two-point function in the IR limit becomes
\be
\bra O ( \ta, \vec{x}_1) O( \ta, \vec{x}_2) \ket  &\sim& e^{-  |\vec{x}_1-\vec{x}_2|  /\xi_{ir}}  ,
\ee
where the interaction with the thermalized LFT excitation gives rise to an effective mass 
\be
m_{eff} \equiv \fr{1}{\xi_{ir}}= \D \ls \fr{   4 \pi }{ d+\lambda -1}  \rs^{1/\l} \
   T_H^{1/\lambda }  ,
\ee
where $\xi_{ir}$ means a correlation length. This result shows that the effective mass in the IR region is proportional to $T_H^{1/\l}$. Therefore, the higher temperature is, the shorter the correlation length becomes.

For a Euclidean Lifshitz black hole with a general $\l$, a temporal geodesic length is given by
 \be
L( \ta_1 , \vec{x} ; \ta_2 ,\vec{x} ) =  \int_0^{| \ta_1-\ta_2| } d \ta  \  \fr{R}{z} \  \sqrt{\fr{R^{2(\l-1)}}{z^{2(\l-1)}} f + \fr{1}{f} \ls \fr{dz}{d \ta} \rs^2} ,    \la{Relation:EGeo}
\ee
where the Euclidean time $\ta$ is periodic. Using the Wick rotation ($\ta = i t$), this Euclidean black hole is rewritten as the Lorentzian one where the Lorentzian time $t$ is not periodic.  To see the relation between Euclidean and Lorentzian two-point functions, we first focus on the case of $\l=1$. Using the conserved quantity, $dz/d \ta$ is given by
\be
\fr{dz}{d \ta} = \fr{f(z)  \sqrt{f z_0^2 - f_0 z^2} }{z \sqrt{f_0}} , \la{Relation:turn}
\ee
where $f_0$ is the value of $f$ at the turning point. Performing this integral, the turning point is determined as a function of the time interval
\be
z_0 = z_h \sin \ls \fr{ | \ta_1 - \ta_2 |}{2 z_h} \rs .   \la{Result:ETP}
\ee
Using \eq{Relation:turn} and \eq{Result:ETP}, the geodesic length reads
\be
L( \ta_1 , \vec{x} ; \ta_2 ,\vec{x} )  = 2 R \log \lb \fr{2 z_h}{\e} \sin \ls \fr{ | \ta_1 - \ta_2 |}{2 z_h} \rs  \rb ,
\ee
and a Euclidean temporal two-point function is given by a periodic function
\be
\bra {\cal O} (\ta_1,x) {\cal O} (\ta_2,x) \ket =  \fr{\e^{2\D}}{2^{2\D} z_h^{2\D}} \fr{1}{ \sin\ls   | \ta_1 - \ta_2 |/(2 z_h)  \rs^{2 \D}} .
\ee
Using the Wick rotation, a Lorentzian temporal two-point function becomes non-periodic
\be
\bra {\cal O} (t_1,x) {\cal O} (t_2,x) \ket =  \fr{\e^{2\D}}{(2 i)^{2\D} z_h^{2\D}} \fr{1}{ \sinh \ls  | t_1 - t_2 |/(2 z_h)  \rs^{2 \D}} .  \la{Result:ET2ptF}
\ee
In the long time interval limit ($| t_1 - t_2 |/z_h \to \infty$), the Lorentzian two-point function exponentially suppresses 
\be
\bra {\cal O} (t_1,x) {\cal O} (t_2,x) \ket \sim    e^{- \D  | t_1 - t_2 |/ z_h} .   \la{Result:lambda1}
\ee

In the long time interval limit, the exponential suppression of  two-point functions generally appears. To see this, we first write a time interval and geodesic length as functions of a turning point
\be
| \ta_1-\ta_2|  &=& \fr{2 \sqrt{f_0}  }{R^{\lambda-1} }  \int_0^{z_0} dz  \frac{  z^{2 \lambda -1}}{f \sqrt{f z_0^{2 \lambda }-f_0 z^{2 \lambda }}} , \la{Relation:Tintval} \\
L ( \ta_1 , \vec{x} ; \ta_2 ,\vec{x} )
&=& \int_0^{z_0} dz  \frac{2 R z_0^{\lambda }}{z \sqrt{f z_0^{2 \lambda }-f_0 z^{2 \lambda }}}   .  \la{Relation:TGoeLength}
\ee
Then, a Euclidean geodesic length \eq{Relation:EGeo} can be rewritten as the following form  
\be
L ( \ta_1 , \vec{x} ; \ta_2 ,\vec{x} )  = \fr{R^\l}{z_0^\l } \,  | \ta_1-\ta_2| 
+ \fr{2 R }{z_0^{\lambda }} \int_0^{z_0} dz \frac{ f z_0^{2 \lambda }-\sqrt{f_0} z^{2 \lambda } }{f z \sqrt{f z_0^{2 \lambda }-f_0 z^{2 \lambda }}}  .  \la{Relation:IRB}
\ee 
After applying the Wick rotation, we take an infinite time interval limit ($| t_1-t_2|  \to \infty$), which appears at $z_0 \to z_h$. In this limit, $f_0 \to 0$ and the second term of the right hand side in \eq{Relation:IRB}  reduces to
\be
\lim_{z_0 \to z_h} \fr{2 R }{z_0^{\lambda }} \int_0^{z_0} dz \frac{ f z_0^{2 \lambda }-\sqrt{f_0} z^{2 \lambda } }{f z \sqrt{f z_0^{2 \lambda }-f_0 z^{2 \lambda }}}  = \int_0^{z_h} dz  \fr{2 R}{\sqrt{f} z}  \la{Result:diffLtau}
\ee
Performing this integral leads to a UV logarithmic divergence at $z=0$. In general, a UV divergence can be removed by an appropriate renormalization scheme. Ignoring the UV divergence, the integral \eq{Result:diffLtau} is finite if $f$ has only a simple root. Therefore, the geodesic length becomes proportional to the time interval. As a result, a temporal two-point function in the infinite time interval limit reduces to
\be
\bra O (t_1, \vec{x}) O( t_2, \vec{x}) \ket \sim e^{- \D R^{\l-1}| t_1-t_2| / z_h^\l  }  .  \la{Result:IRtemp}
\ee
For $\l=1$, this is consistent with the previous CFT result in \eq{Result:lambda1}. This IR temporal 
two-point function determines a relaxation or half-life time of the operator, which interacts with the LFT excitation, as
\be
t_{1/2} 
=  \frac{(d+\l-1)}{4 \pi  \D R^{\l-1}  }  \fr{1}{  T_H   }  .
\ee
This shows that the temporal two-point function of a thermal LFT exponentially suppresses with a half-life time inversely proportional to temperature.

\section{ RG flow of two-point functions in a disorder system}

Recently, the entanglement entropy of a randomly disordered system was studied in the holographic setup \cite{Hartnoll:2014cua,Arean:2013mta,Narayanan:2018ilr}. The disorder allows a UV CFT to flow into a new IR LFT. In the holographic setup, the disorder is described by the following Euclidean Einstein-scalar gravity
\be
S = \fr{1}{16 \pi G} \int d^3 x \sqrt{g} \ls {\cal R} - \fr{2}{R^2} + 2 \pa_a \ph \pa^a \ph + 2m^2 \ph^2 \rs ,
\ee
where the scalar field represents the disorder and we take $m^2  =-3/ (4 R^2) $. For convenience, we set $R=1$ from now on. The scalar field in an asymptotic AdS allows the following perturbative expansion
\be
\ph = \ph_1 (x) \,  u^{1/2} + \ph_2 (x)  \, u^{3/2} + \cdots ,
\ee
where $u$ indicates the radial coordinate. The the scalar field rapidly suppresses in the asymptotic region.

To describe the random disorder, we assume that $\ph_1$ is given by (see \cite{Hartnoll:2014cua} for more details)
\be
\ph_1 (x) = v \sum_{n=1}^{N-1} A_n \cos\ls k_n x + \g_n \rs ,
\ee
where $k_n=n \D k$ and $\D k =k_0 /N$ with a highest mode $k_0$. Here, $\g_n$ denotes a random phase and the disorder amplitude is given by $A_n = 2 \sqrt{S(k_n) \ \D k}$ where $S(k_n)$ represents correlation of the noise. When $S(k_n)=1$, the disorder leads to the Gaussian distribution
\be
\bra \ph_1 (x) \ket_R = 0 ~~ {\rm and} ~~ \bra \ph_1 (x) \ph_1 (y)\ket_R = v^2 \d(x-y) .
\ee
The gravitational backreaction of this scalar field determines the dual geometry up to $v^2$ order \cite{Hartnoll:2014cua}
\be
ds^2 = \fr{1}{u^2} \ls  \fr{A(u)}{F(u)^{v^2/2}} d \ta^2 + B(u) dx^2 + du^2 \rs ,    \la{Metric:DisdeformedAdS}
\ee
where
\be
A(u) &=& 1  + v^2 \lb e^{- 2 k_0 u} \lc (1+2 k_0 u) \ls \,\log (2k_0 u)+\b \rs  -   2 k_0 u  \fr{}{}\rc   - \log \ls \fr{2 k_0 u}{\sqrt{1  + k_0^2 u^2} } \rs \rb , \nn
B(u) &=& 1 + v^2 \ls e^{- 2 k_0 u} +\b -1 \rs , \nn
F(u) &=& 1+k_0^2 u^2  .   \la{Solution:dualLFT}
\ee
Here $\b$ is the Euler constant, $\b \approx 0.577$. This geometric solution shows that an asymptotic AdS geometry smoothly changes into a Lifshitz one as $u$ increases. In this case, the dynamical critical exponent continuously changes from $\l=1$ in the UV limit into $\l = 1+ {v^2}/{2}$ in the IR limit.

To define two-point functions consistently with a disorder deformation, we first consider an asymptotic geometry of \eq{Metric:DisdeformedAdS}. In the UV limit ($u \to 0$), the metric reduces to
\be
ds^2_{UV} = \fr{B_{UV}}{u^2}   \ls  d \ta^2 + dx^2 \rs +  \fr{du^2 }{u^2}  ,
\ee
where a constant $B_{UV}$ is given by
\be
B_{UV} = 1 + v^2 \b .
\ee
If we absorb the constant $B_{UV}$ by the coordinate, the above asymptotic metric reduces to an AdS one. In this UV limit, the geodesic length is given by
\be
L_{UV} (|x_1 - x_2| ) &=& \int^{ |x_1 - x_2 |}_{0} dx \ \fr{\sqrt{ B_{UV} + u'^2}}{u} \nn
 &=& 2 \log \ls | x_1 - x_2 |  \rs + \log \ls \fr{1 + v^2 \b}{\e^2} \rs  .
\ee
Introducing a normalized operator ${\cal O} (x_i)$ 
\be
{\cal O} (x_i) = {\cal N} \, O (x_i) ,
\ee
with the following normalization factor at the UV fixed point
\be
{\cal N} = \ls \fr{1 + v^2 \b}{\e^2} \rs^{\D/2} ,
\ee
the normalized UV two-point function becomes
\be
\bra {\cal O} (\ta, x_1) {\cal O} (\ta, x_2) \ket 
=  \fr{1}{| x_1 - x_2 |^{2 \D}} ,		\la{Result:UVtwoptfun}
\ee 
where $\D$ corresponds to the scaling dimension at the UV fixed point. This is the normalized two-point function of a UV CFT.

In general, a scaling dimension is well defined only at fixed points. When a relevant disorder deforms a UV CFT, it makes a nontrivial RG flow. Since the scaling symmetry is broken at intermediate energy scale, the scaling dimension of an operator usually changes along the RG flow. To see more details, we introduce an effective scaling dimension $\D_{eff} (|x_1 - x_2|)$ to specify the scale dependence of two-point functions
\be
\bra {\cal O} (\ta , x_1) {\cal O} (\ta , x_2) \ket =  \fr{1}{| x_1 - x_2 |^{2 \D_{eff} (|x_1 - x_2|)}} .
\ee 
Although the scaling dimension at fixed points is given by a constant, the effective scaling dimension we defined above crucially relies on the energy scale. If we know a two-point function, we can determine the effective scaling dimension by
\be
\D_{eff} (|x_1 - x_2|) = - \half    \fr{d \log \bra {\cal O} (x_1) {\cal O} (x_2) \ket }{d \log |x_1 - x_2|} .
\ee

By using the effective scaling dimension, we investigate how the scaling dimension changes in a disorder system. On the dual geometry of a disorder system, the general geodesic length is governed by
\be
L (| x_1 - x_2 |) = \int_{0}^{| x_1 - x_2 |} dx \ \fr{\sqrt{ B(u) + u'^2}}{u}  ,
\ee
and the normalized spatial two-point function is given by
\be
\bra {\cal O} (\ta, x_1) {\cal O} (\ta, x_2) \ket = {\cal N}^2  \, e^{- \D L(|x_1 - x_2|)  }  ,  \la{Formula:STwoF}
\ee 
where ${\cal N}$ is the normalization factor \eq{Result:UVtwoptfun} which was used to normalize the UV two-point function. After numerically calculating the spatial two-point function in \eq{Formula:STwoF}, we plot the effective scaling dimensions as a function of the operator's distance in Fig. 1(a). The numerical result shows that the disorder reduces the scaling dimension in the short-distance region. In the long-distance limit, on the contrary, the scaling dimension increases and finally approaches the UV scaling dimension at the IR fixed point. This is because the UV CFT and IR LFT have the same scaling behavior in the spatial direction.

\begin{figure}
\begin{center}
\vspace{-0.5cm}
\hspace{-0.3cm}
\subfigure[spatial two-point function]{\label{fig1} \includegraphics[angle=0,width=0.5\textwidth]{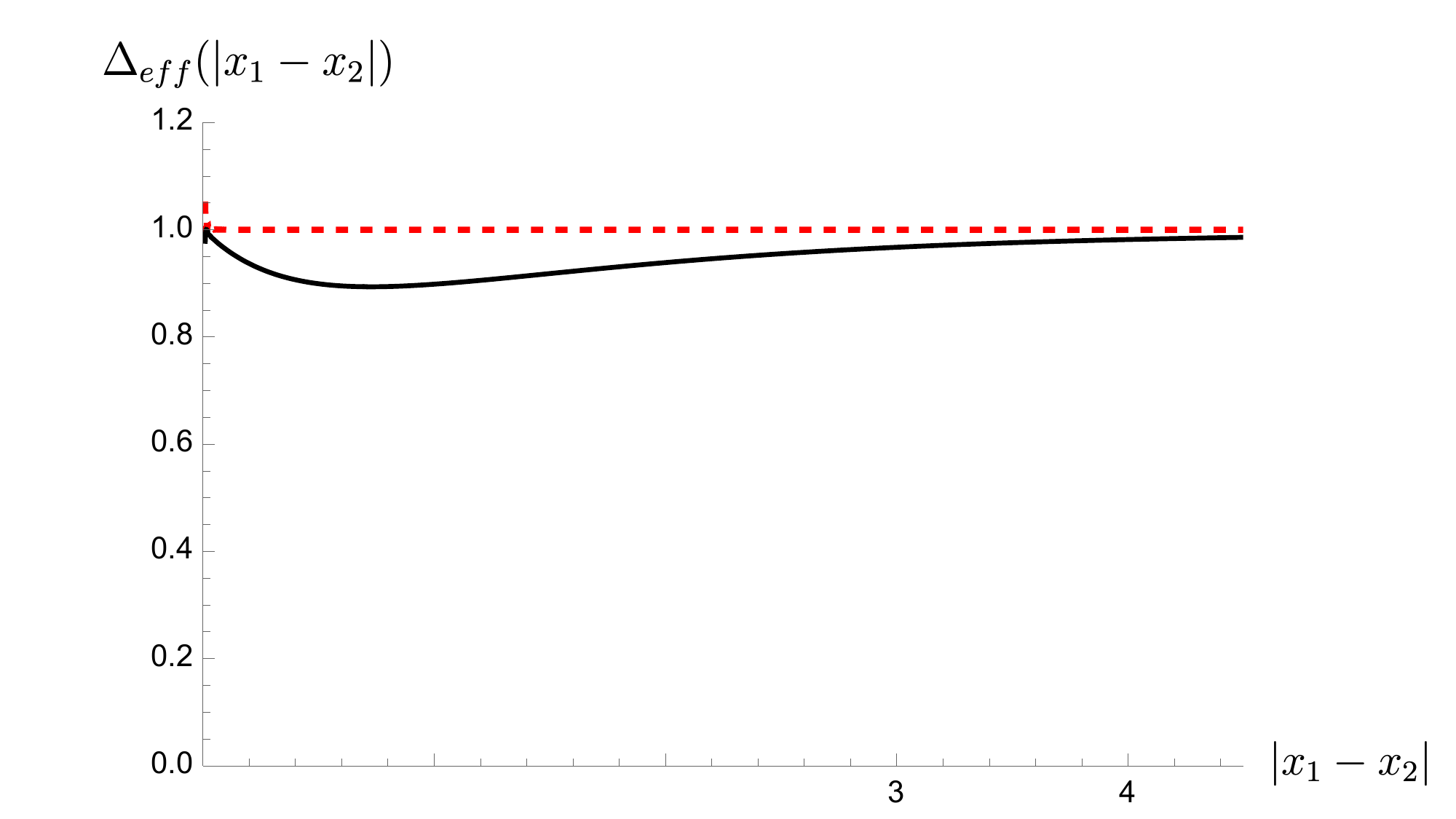}}
\hspace{-0.5cm}
\subfigure[temporal two-point function]{\label{fig2} \includegraphics[angle=0,width=0.5\textwidth]{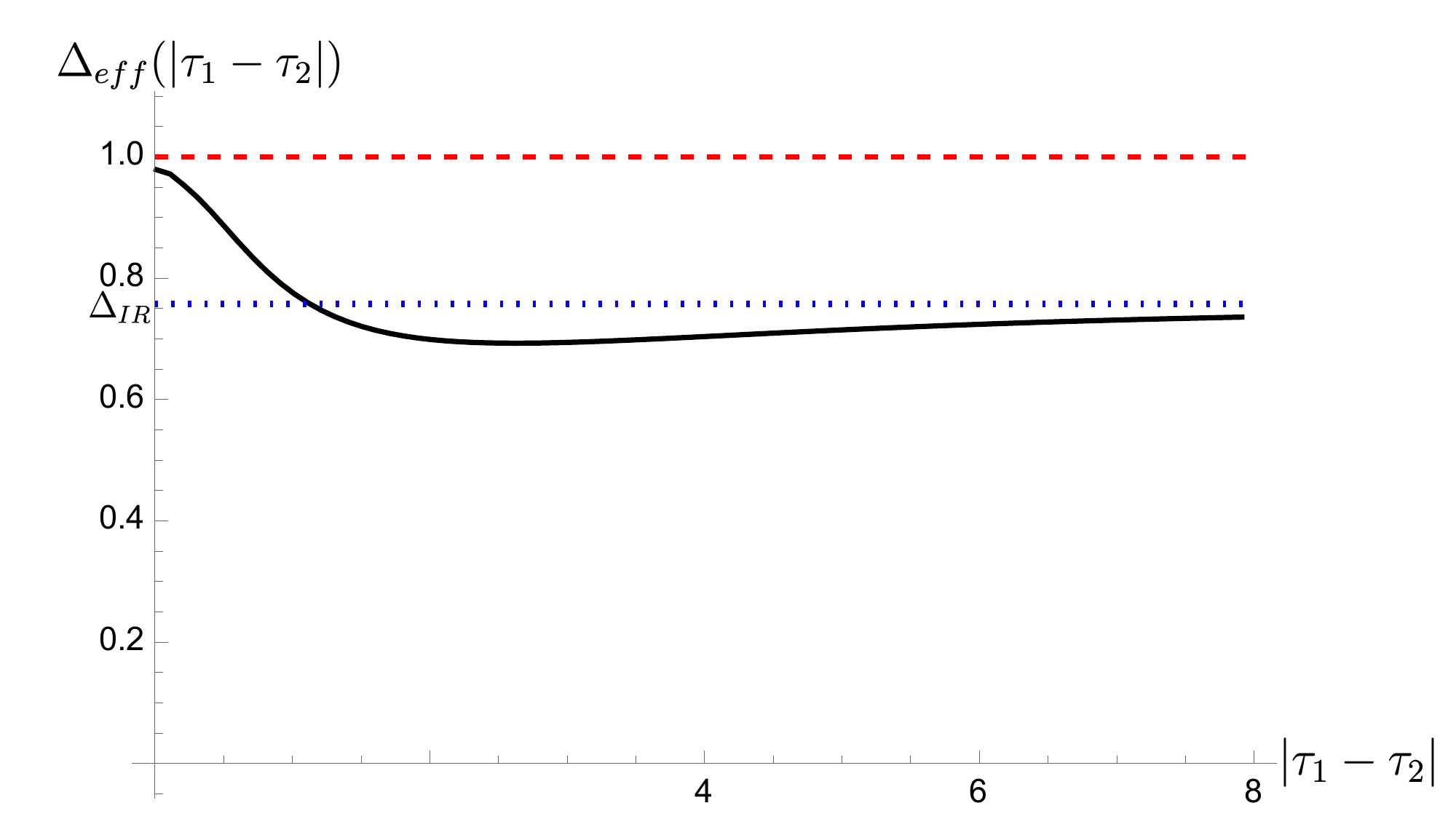}} 
\vspace{-0.cm}
\caption{\small  We depict how the disorder modifies the scaling dimension (black solid curve) for spatial and temporal two-point functions. We take $R=1$, $\D=1$, $k_0 = 2$, and $v=0.8$.  (a) For the spatial two-point function, the scaling dimensions at UV and IR fixed points have the same value, $\D=1$. This is because the same scaling symmetry in the spatial direction occurs at UV and IR fixed points. (b) For the temporal two-point function, the UV scaling dimension $\D=1$ continuously changes into a new IR scaling dimension, $\D_{IR} \approx 0.758$,  after infinite time, which is the value expected in the LFT.}
\label{fig1}
\end{center}
\end{figure}

Similarly, we can also evaluate a temporal two-point function affected by the disorder. Recalling that for $u \to 0$, $A(u)$ and $F(u)$ in \eq{Solution:dualLFT} reduces to $A_{UV} = 1 + v^2 \b$ and $F_{UV} = 1$, the geodesic length for a short time interval is governed by
\be
L_{UV} (| \ta_1 - \ta_2 | )  = \int_{\ta_1}^{\ta_2} d \ta  \ \fr{1}{u} \sqrt{ \fr{A_{UV}}{F_{UV}^{v^2/2}} + \dot{u}^2 }      .
\ee
In the UV limit having a very short time interval ($| \ta_1 - \ta_2 |  \to 0$), the geodesic length is given by
\be
L_{UV} ( | \ta_1 - \ta_2 | ) =  2 \log \ls  | \ta_1 - \ta_2 | \rs + \log \ls \fr{1 + v^2 \b}{\e^2} \rs  .
\ee
In terms of the normalized operator ${\cal O}  (x_i) = {\cal N} O (x_i)$, 
the temporal two-point function of the normalized operator becomes in the UV limit 
\be
\bra {\cal O} (\ta_1,x) {\cal O} (\ta_2,x) \ket 
=  \fr{1}{| \ta_1 - \ta_2 |^{2 \D}} ,
\ee 
which again corresponds to the temporal two-point function of a CFT.


As the time interval becomes large, the disorder affects the temporal two-point function and modifies the operator's scaling dimension. When we consider a geodesic length extending to the deformed geometry  \eq{Metric:DisdeformedAdS}
\be
L (| \ta_1 - \ta_2 | )  = \int_{\ta_1}^{\ta_2} d \ta  \ \fr{1}{u} \sqrt{ \fr{A(u)}{F(u)^{v^2/2}}+ \dot{u}^2 }     ,
\ee
the temporal two-point function of the normalized operator is given by
\be
\bra {\cal O} (\ta_1, x) {\cal O} (\ta_2,x) \ket 
=  \fr{1}{| \ta_1 - \ta_2 |^{2 \D_{eff} (|\ta_1 - \ta_2|)}} .
\ee 
After numerical calculation, we depict how the disorder modifies the scaling dimension of an operator in Fig. 1(b).   Fig. 1(b) shows that the effective scaling dimension gradually decreases in a short time interval. After the critical time where the scaling dimension has a minimum value,  the scaling dimension increases with time. After infinite time, the scaling dimension finally approaches a constant value, $\D_{IR} \approx 0.758 $, which is the value expected by the IR LFT for $\D=1$ and $v=0.8$
\be
\D_{IR} = \lim_{|\ta_1 - \ta_2| \to \infty}  \D_{eff} (|\ta_1 - \ta_2|) = \fr{\D}{1+ v^2/2}  . 	\la{Result:IReffdim}
\ee
Since the disorder breaks the Lorentz symmetry, the spatial and temporal two-point functions at the IR fixed point lead to different scaling dimensions which are expected in the LFT. 



\section{Discussion}

We investigated the holographic two-point functions of nonconformal field theories and their RG flow. In the holography study, some physical quantities of QFTs can be represented as geometrical objects in the dual gravity. They are useful to understand nonperturbative features of strongly interacting QFTs. One of them is a geodesic length connecting two boundary operators, which corresponds to a two-point function of a dual QFT. Although there is no rigorous proof for this connection, it is manifest for CFTs because of the large symmetry. We showed that this is also true for nonconformal field theories. To do so, we took into account a LFT which may appear at a critical or IR fixed point. After explicitly calculating a geodesic length in the Lifshitz geometry, we explicitly showed that this holographic calculation reproduces the known two-point functions of the LFT vacuum. Using the same holographic method, we further studied two-point functions of an operator interacting with the LFT excitation. In the UV region, the screening effect of the background LFT excitation slightly modifies the two-point function by reducing the strength of correlation. In this case, the first correction is proportional to the energy of the LFT excitation. This feature becomes more manifest in the IR regime. We showed that two-point functions involving the interaction with the LFT excitation suppresses exponentially in the IR limit. For the temporal two-point function, in particular, such suppression determines a half-life time, which is inversely proportional to temperature.

Lastly, we looked into how a relevant disorder affects the scaling dimension of an operator when a UV CFT flows into a LFT. In general, a nonconformal field theory has nontrivial two-point functions including all quantum corrections. Although it is important to understand such quantum feature, it is generally hard to calculate all quantum corrections exactly if there is no sufficiently large symmetry. However, we showed that the holographic calculation allows us to investigate how a relevant disorder modifies the scaling dimension of an operator. When a disorder changes a UV CFT into an IR LFT, we calculated the two-point functions and described the change of the scaling dimension along the RG flow. On the dual gravity side, we reproduced the expected LFT's two-point functions in the IR limit from UV CFT's ones. Using the results studied in this work, it would be interesting to figure out an anomalous dimension and Callan-Symannzik equation from the effective scaling dimension defined in this work. In future works, we hope to report more interesting results on this issue.

\vspace{0.5cm}

{\bf Acknowledgement}

This work was supported by the National Research Foundation of Korea(NRF) grant funded by the Korea government(MSIT) (No. NRF-2019R1A2C1006639).




%

\end{document}